\documentclass[
    ,final            
  ]
  {aipproc}

\layoutstyle{6x9}

\begin{document}

\title{An Integrated Tracker for STAR}


\author{Frank Simon (for the STAR collaboration)}{
  address={Massachusetts Institute of Technology}
}

\begin{abstract}
The STAR experiment at the Relativistic Heavy Ion Collider RHIC studies the new state of matter produced in relativistic heavy ion collisions and the spin structure of the nucleon in collisions of polarized protons. In order to improve the capabilities for heavy flavor measurements and the reconstruction of charged vector bosons an upgrade of the tracking system both in the central and the forward region is pursued. The integrated system providing high resolution tracking and secondary vertex reconstruction capabilities will use silicon pixel, strip and GEM technology.

\end{abstract}

\maketitle


\section{Introduction and Current Capabilities}

The STAR experiment at RHIC studies the fundamental properties of the new state of strongly interacting matter produced in relativistic heavy ion collisions and investigates the spin structure of the proton in polarized $p+p$ collisions. A variety of results both in heavy ion collisions and polarized $p+p$ collisions have already been obtained. A key future step in these programs is the ability for direct reconstruction of particles containing charm and bottom quarks as well as flavor tagging of jets to allow precise measurements of the spectra, yields and flow of open charm and bottom and to determine spin dependent production asymmetries connected to the gluon polarization in the nucleon. The flavor dependence of the sea quark polarization will be determined by parity violating W production and decay in longitudinally polarized $p+p$ collisions at $\sqrt{s}$ = 500 GeV. 

STAR \cite{Ackermann:2002ad} is one of the two large detector systems at RHIC. Its main tracking detector is a large-volume time projection chamber (TPC) covering the pseudorapidity range $\vert \eta \vert < 1.2$. Additional vertex resolution for the reconstruction of secondary decay vertices is provided by the silicon vertex tracker (SVT, $\vert \eta \vert < 1$), a three--layer silicon drift detector, and the one--layer silicon strip detector (SSD). Tracking in the forward region is provided by the forward TPCs (FTPCs, $2.5 < \vert \eta \vert < 4.0$). The barrel (BEMC) and endcap (EEMC) electromagnetic calorimeters cover $-1 < \eta < 1$ and $1 < \eta < 2$, respectively. Additional small acceptance electromagnetic calorimetry at high rapidity is provided by the forward pion detector (FPD, $3.1 < \vert \eta \vert < 4.2$). 

\begin{figure}
\includegraphics[width = 0.95\textwidth]{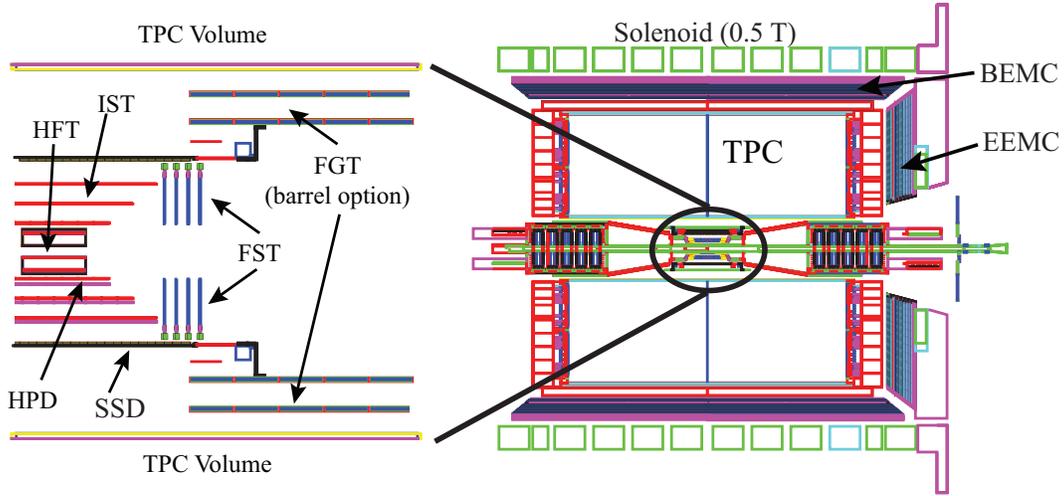}
\caption{Side view of the STAR detector with planned tracking upgrades. The inner tracking region is shown enlarged. The inner tracking system covering $\vert \eta \vert < 1$ consists of the HFT, HPD the IST and the existing SSD. The forward tracking system covering $ 1< \eta < 2$ consists of the FST and the FGT. In addition to the barrel layout for the FGT shown here a disk option is also being investigated.}
\label{fig:TUP}
\end{figure}

The current tracking capabilities are insufficient to address the future measurements outlined above. The planned integrated tracker is designed to provide the necessary vertex resolution to uniquely identify open charm and bottom and to provide precision tracking in the forward region to determine the charge sign of electrons from $W^+$ and $W^-$ decays that are detected in the EEMC. Figure \ref{fig:TUP} shows an overview of the planned tracking upgrades for STAR. The two distinct areas of inner and forward tracking are driven by different physics motivations, outlined in the following sections together with the technology choices for the planned upgrades.

\section{Inner Tracker} 

Heavy quarks are good probes for the properties of the matter created in relativistic heavy ion collisions \cite{Adams:2005dq}. Due to their high intrinsic mass, frequent interactions are needed to bring $c$ and $b$ quarks into equilibrium with the surrounding matter. Collective flow of heavy quarks is thus a strong indication of thermalization in the early stages of the reaction. Flavor tagged jets will provide information on the energy loss of light versus heavy quarks in the created medium. 
The production of heavy quarks in $p+p$ collisions is dominated by gluon-gluon fusion, $gg \rightarrow c\bar{c}, b\bar{b}$. The double longitudinal spin asymmetry in this process thus provides direct access to the gluon polarization in the proton and is largely independent of the quark helicity distributions \cite{Bunce:2000uv}.

Since $c\tau\, \sim 120\ \mu\mbox{m}$ for $D^0$ and $c\tau\, \sim 460\ \mu\mbox{m}$ for $B^0$ excellent vertex resolution is needed to directly identify these particles. The planned upgrade for the STAR inner tracker is designed to achieve this both in heavy ion collisions and in polarized $p+p$ collisions by an optimization for high multiplicity and high rate environments. A thin beryllium beam pipe with a radius of 2 cm will be used to give the detectors close access to the collision point. The inner tracker consists of three devices, all covering $\vert \eta \vert$ < 1.0. The Heavy Flavor Tracker HFT \cite{Xu:2006dx} is a lightweight two--layer detector based on Active Pixel Sensors (APS) with 30 $\mu$m $\times$ 30 $\mu$m pixels using silicon thinned down to 50 $\mu$m, limiting the material of the detector to $\sim$ 0.3\% $X_0$ per layer. The inner sensor layer sits at a radius of 2.5 cm and a staggered outer layer sits at 6.5 cm and 7.5 cm radius. The radius was increased with respect to the numbers in the original proposal due to an increased radius of the beam pipe in current plans. This device will provide a spatial resolution better than 10 $\mu$m at the inner layer. A fast intermediate tracker is needed to act as a pointing device from the TPC to the HFT to connect the precision points in the HFT to TPC tracks and to provide the time resolution necessary for high luminosity running. Outside the HFT a single--layer hybrid pixel detector (HPD), using readout cells of 50 $\mu$m $\times$ 425 $\mu$m, is planned at a radius of 9.1 cm. It is based on the ALICE silicon pixel detector \cite{Riedler:2005vr}, using the same chips and mechanical structure, albeit with slightly longer ladders to provide for the larger radius. The gap to the existing SSD at 23 cm will be bridged by the intermediate silicon tracker IST, consisting of two layers of conventional back-to-back silicon strip sensors at 12 cm and at 17 cm radius. The material budget for this fast device is estimated to be $\sim$ 1.5 \% $X_0$ per layer, similar to that of the existing SVT. The intermediate tracker will replace the SVT which does not have sufficient rate capability for future collider luminosities and is incompatible with future upgrades of the STAR data acquisition system. The precise layout, the number of layers and the technology choices for the intermediate tracker are currently being studied in simulations aimed at an optimized design for the inner tracker. 

\vspace*{-3mm}
\section{Forward Tracker}

From polarized deep inelastic scattering experiments it is known that the flavor-integrated contribution of quarks to the proton spin is surprisingly small. A measurement of the polarization of the quark sea through a flavor separated study of quark and anti-quark polarizations is thus of fundamental interest \cite{Bunce:2000uv}. At RHIC, flavor separated measurements will be carried out via the maximally parity violating production of $W$ bosons in $u\bar{d} \rightarrow W^+$ and $d\bar{u} \rightarrow W^-$ reactions. These reactions are ideal to access the quark polarizations since the $W$ boson couples only to left-handed quarks and right-handed anti-quarks. For $W$ production away from mid-rapidity, the quark is most likely a valence quark from the proton traveling in the same direction as the produced $W$, while the anti-quark comes from the sea of the other proton. That way the spin state of the proton is cleanly linked to the partons involved in the reaction. 

\begin{figure}
\begin{minipage}{0.495\textwidth}
\includegraphics[width=\textwidth]{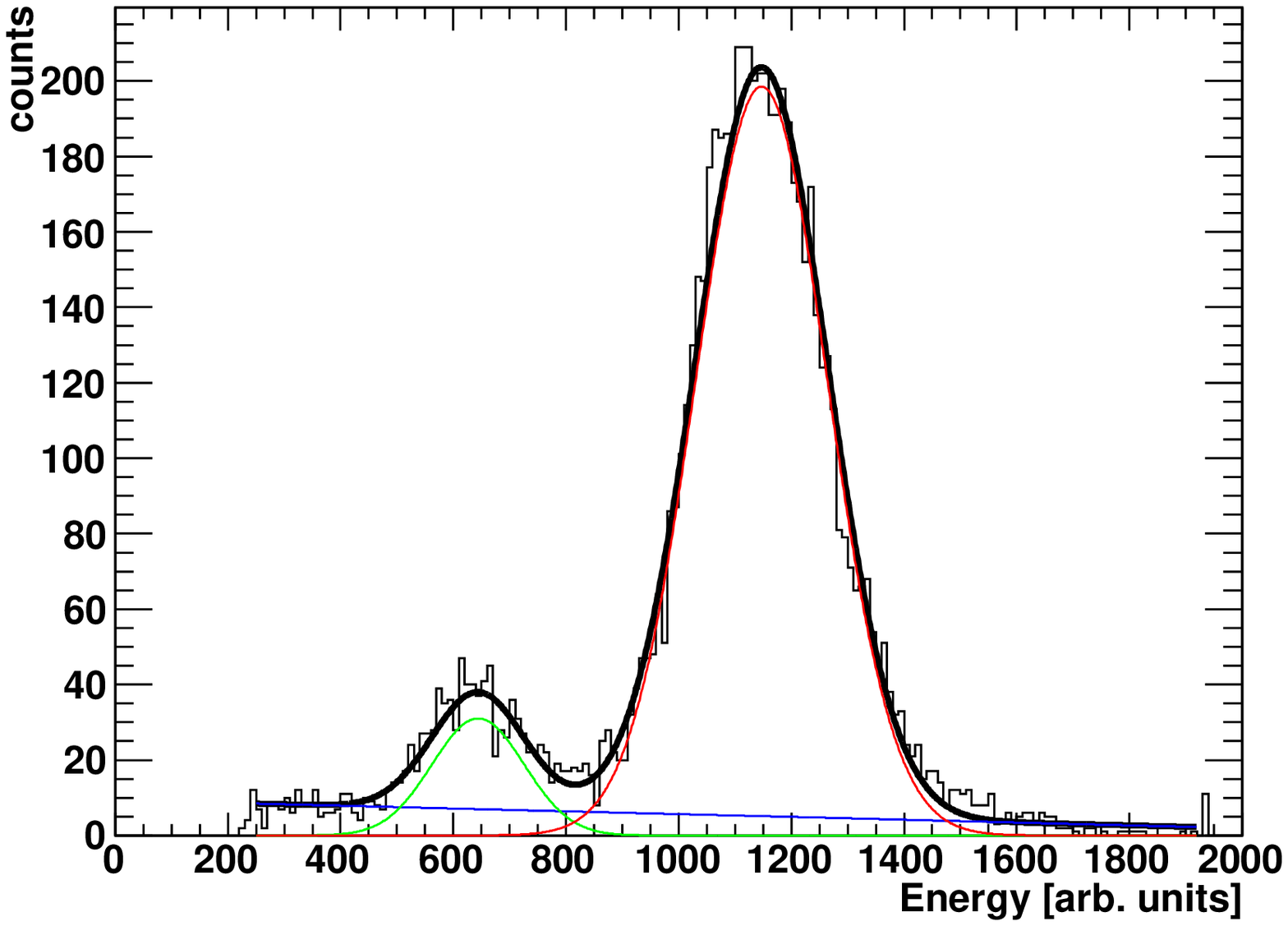}
\end{minipage}
\begin{minipage}{0.495\textwidth}
\includegraphics[width = \textwidth]{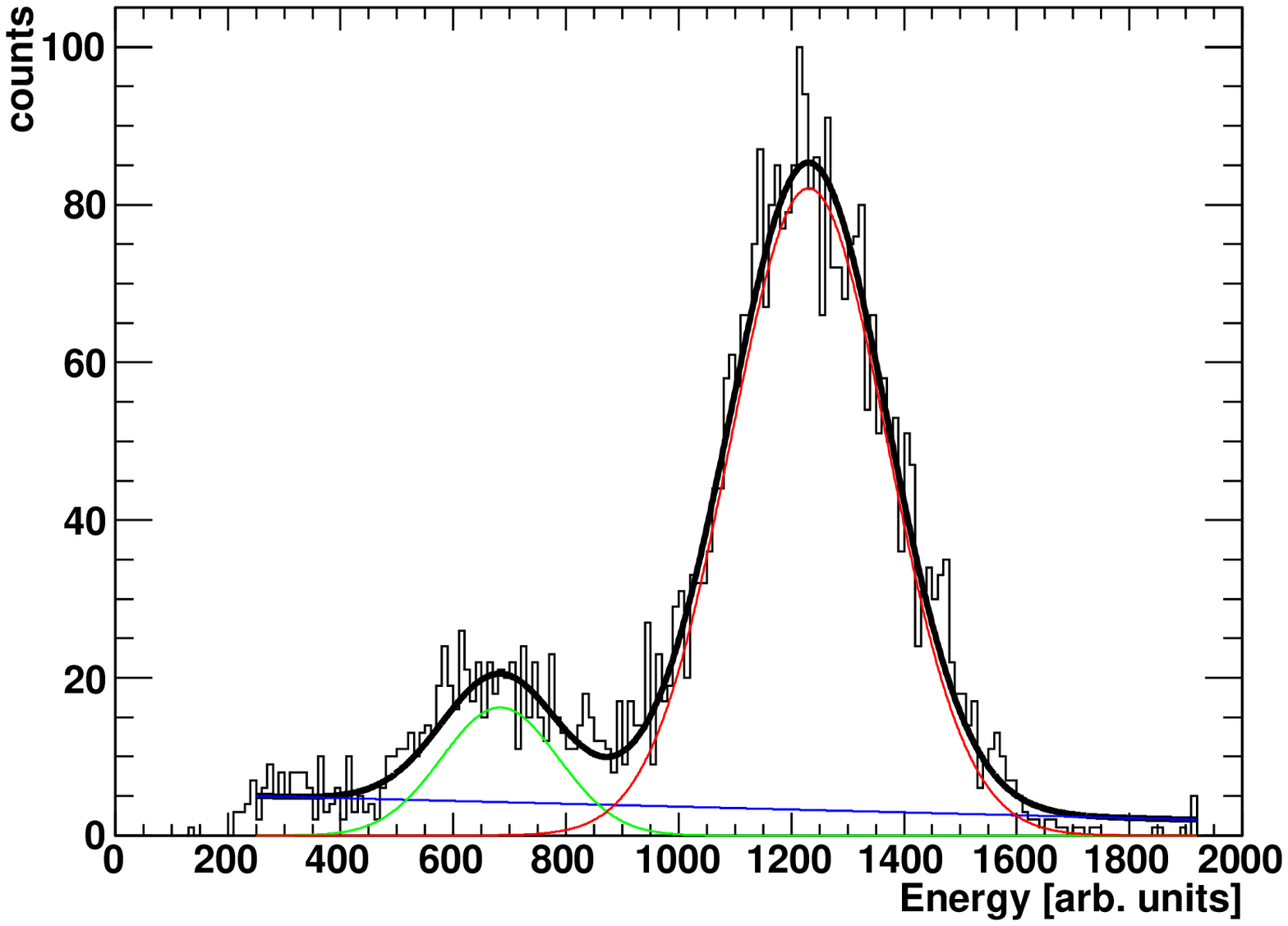}
\end{minipage}
\caption{Typical $^{55}$Fe X-ray spectra taken with triple GEM test detectors using foils manufactured at CERN (left) and at TechEtch (right). The spectra are fitted with the sum of two Gaussians and a linear background. The energy resolution (FWHM of the photopeak divided by the mean) is on the order of 20\% for both detectors.}
\label{fig:Spectrum}
\end{figure}

At STAR, produced $W$s will be detected via their leptonic decays into an electron and a neutrino, $W^+ \rightarrow e^+ \nu_e$ and $W^- \rightarrow e^- \bar{\nu}_e$. The energy of the forward going lepton will be measured in the EEMC, providing a clean signature for a $W$ decay. It is crucial to distinguish between $W^+$ and $W^-$ since this carries the information on the flavor of the colliding quarks. This is achieved by identifying the charge sign of the high momentum lepton from the $W$ decay, requiring  high resolution tracking in the acceptance of the EEMC from 1 to 2 in $\eta$, a region currently not covered by trackers in STAR. The necessary resolution will be provided by two detector systems. The Forward Silicon Tracker FST will consist of up to 4 silicon disks using conventional back-to-back silicon strip detectors close to the interaction point. The Forward GEM Tracker FGT will provide additional space points with a larger lever arm. Two geometries are currently being evaluated for this device, namely a two layer barrel (each layer providing a space point) and an option with multiple discs along the beam axis that provide at least two points on a track for $1 < \eta < 2$. 
The FGT will be based on GEM technology \cite{Sauli:1997qp}, using a triple GEM configuration similar to the one successfully applied by the COMPASS experiment \cite{Altunbas:2002ds}. The front-end electronics for this detector will be based on the APV25-S1 chip \cite{French:2001xb}, which will also be used for the FST and the IST, significantly reducing development costs for the readout and data acquisition system. For this large-scale project the commercial availability of GEM foils is necessary. A collaboration with TechEtch Inc. of Plymouth, MA, USA has been established to develop the production process for these foils. Figure \ref{fig:Spectrum} shows typical $^{55}$Fe X-ray spectra (main line at 5.9 keV) recorded with triple GEM test detectors using CERN and TechEtch made foils. The test detectors are read out via standard preamplifier and amplifier setups, collecting the full charge in a single channel. The energy resolution and signal quality is comparable for CERN and TechEtch made foils. Some issues with gain stability over time still exist with the TechEtch foils which are currently being investigated. 

\vspace*{-3mm}
\section{Summary}

The STAR collaboration is preparing a challenging tracking upgrade program to further investigate the properties of the new state of strongly interacting matter produced in relativistic heavy ion collisions and to provide fundamental studies of the nucleon spin structure in high-energy polarized proton-proton collisions. Key elements are the ability to directly reconstruct charm and bottom decays and to determine the charge sign of electrons produced in $W$ decays. The mid-rapidity inner tracker includes high resolution active silicon pixel sensors, hybrid pixels and standard single sided silicon strip detectors. The forward tracker is based on silicon strip discs and large area triple-GEM trackers. 


\vspace*{-3mm}

\bibliographystyle{aipproc}   
\bibliography{FSimonSTARTracking}

\end{document}